\documentclass[aps,amssymb,twocolumn,unsortedaddress]{revtex4}
\usepackage{setspace}
\usepackage{graphicx}
\usepackage{psfrag}

\newcommand{\ps}{p^\star}
\newcommand{\taus}{\tau^\star}
\newcommand{\brho}{\beta^\rho}
\newcommand{\bz}{\beta^z}
\newcommand{\sigmabar}{\bar{\sigma}}
\newcommand{\omegabar}{\bar{\Omega}}
\newcommand{\pbar}{\bar{p}}

\begin{document}
\title{Critical Collapse of the Massless Scalar Field in Axisymmetry}

\author{Matthew W. Choptuik}
\affiliation{CIAR Cosmology and Gravity Program \\
     Department of Physics and Astronomy,
     University of British Columbia,
     Vancouver BC, V6T 1Z1 Canada}
\author{Eric W. Hirschmann}
\affiliation{Department of Physics and Astronomy,
     Brigham Young University,
     Provo, UT 84604}
\author{Steven L. Liebling}
\affiliation{Southampton College, Long Island University,
     Southampton, NY 11968}
\author{Frans Pretorius}
\affiliation{Theoretical Astrophysics,
     California Institute of Technology,
     Pasadena, CA 91125}

\begin{abstract}
We present results from a numerical study of critical gravitational
collapse of axisymmetric distributions of massless scalar field energy. 
We find threshold behavior that can be described by the spherically 
symmetric critical solution with axisymmetric perturbations. 
However, we see indications of a growing, non-spherical mode
about the spherically symmetric critical solution.
The effect of this instability is that the small asymmetry
present in what would otherwise be a spherically symmetric self-similar
solution grows. This growth continues until a bifurcation 
occurs and two distinct regions form on the axis, each resembling
the spherically symmetric  self-similar solution. The existence 
of a non-spherical unstable mode is in conflict with previous perturbative 
results, and we therefore discuss whether 
such a mode exists in the continuum limit, or whether we are instead 
seeing a marginally
stable mode that is rendered unstable by numerical approximation.
\end{abstract}

\maketitle

\section{Introduction}\label{intro}

In this paper we present results from a numerical study of critical
collapse of the massless scalar field in axisymmetry.
In spherical symmetry, the threshold of black hole formation was first 
systematically explored in~\cite{choptuik}, which described intriguing
behavior, called {\em critical phenomena}, in solutions approaching the 
threshold. This behavior includes {\em power-law scaling} of the mass
$M$ of black holes that form in the super-critical regime,
\begin{equation}
M \propto \left(p-\ps\right)^\gamma,
\end{equation}
where $\gamma$ is a {\em universal} constant (i.e. independent of the 
initial data) called the {\em scaling exponent}. Here, 
$p$ is a parameter describing some aspect of the initial distribution
of scalar field energy such that for $p>\ps$ black holes form during
evolution, while for $p<\ps$ all of the scalar field disperses to infinity.
Thus, $\ps$ denotes the threshold of black hole formation for the particular 
family of initial data under consideration.
The solution approached in the limit $p\rightarrow \ps$, called the 
critical solution, was also conjectured to be universal, in that
all one-parameter ($p$) families of initial data having a threshold
parameter $\ps$ should exhibit the {\em same} critical solution in the
vicinity of collapse. In addition, the critical solution for the real
scalar field is discretely self-similar,
characterized by an {\em echoing exponent} $\Delta$.
Since the initial discovery reported in~\cite{choptuik}, critical phenomena 
have been observed
in numerous systems---see \cite{gundlach,wang} for recent review articles on the
subject. Note that the particular behavior observed in the threshold
solution depends upon the matter model and spacetime dimensionality.

To date, the only non-perturbative calculation of critical gravitational
collapse away from spherical symmetry was carried out by Abrahams and Evans 
\cite{abrahams_evans}, who studied the collapse of pure gravitational
waves in axisymmetry (note that
axisymmetry is the `minimal' symmetry one can impose on gravitational
waves and retain the possibility of black hole formation). 
In addition, the threshold of singularity formation
in a non-linear sigma model in three dimensions was considered 
in~\cite{liebling}, and found to exhibit features similar to critical
gravitational collapse. These studies provide evidence that critical
phenomena {\em is} observed beyond spherical symmetry at the respective 
thresholds of these two distinct physical systems.

An explanation for critical phenomena, in particular the observed
universality of the solution and departures from it in near-critical
collapse, is offered by positing that the critical solution, when
perturbed, has {\em exactly one unstable mode} \cite{unstable_mode}.
That there is only one unstable mode allows the threshold
solution to be found in a numerical collapse ``experiment'' whereby we
fine-tune a single parameter of a generic family of initial data.
Furthermore, the nature of the unstable
mode eventually dominates the properties of near-critical solutions; for
example, the scaling exponent $\gamma$ can be shown to be
equal to the inverse of the exponential growth factor $\lambda$
of the unstable mode.

The purpose of the present study is to move beyond spherical symmetry
and to explore the threshold of 
black hole formation from the collapse of axisymmetric distributions
of the massless scalar field. 
Linear perturbation studies of the scalar field critical solution 
beyond spherical symmetry were carried out by Mart\'{i}n-Garc\'{i}a and 
Gundlach \cite{martin_garcia_gundlach} (a similar analysis has also been
performed by Gundlach \cite{gundlach2} for the case of perfect fluid
collapse). Their study found no additional growing modes beyond the 
one seen in spherical collapse, a result that suggests
that we should expect to see the spherical critical solution
emerge from our axisymmetric studies.

Having looked at a variety of initial
configurations of the scalar field, some deviating significantly from 
spherical symmetry, we find that in all cases during the early 
phases of near-critical
evolution, we {\em do} see a discretely self-similar solution unfold
that can be described as the spherically symmetric critical solution
plus perturbations. However, in contrast to the perturbation theory
calculations in \cite{martin_garcia_gundlach}, we find some evidence
for a second, slowly growing unstable mode, with an angular dependence 
described by the $\ell=2$ spherical harmonic. The 
simulations suggest that this mode will eventually cause a near-critical
solution, with some asymmetries, to ``bifurcate'' into two distinct
echoing solutions, which, individually, would subsequently be subject 
to the same instability.
In principle then, if we could fine tune to 
arbitrary precision, this bifurcate behavior would be repeated indefinitely.

The appearance of this second unstable mode is in conflict
with the above-mentioned perturbative results. One possibility is
that the non-spherical mode that appears unstable in our simulations
is in fact damped in the continuum
limit, and only grows within the context of our discrete  numerical 
approximation.
Our current code (running on the computer systems
to which we have access) cannot provide the accuracy needed to 
{\em conclusively} 
determine that the growth rate of the suspect mode is positive in 
the continuum limit, and not dominated by truncation error effects. 

The remainder of this paper is organized as follows. In Sec. \ref{system}
we briefly describe the relevant system of equations, the numerical
code used to solve them, and various properties of the
solution that we will analyze. Details of the formalism and numerical
technique can be found in \cite{graxi_intro}. 
In Sec. \ref{results} we describe several
of the families of initial data that we have studied, and present the
results from corresponding near-critical collapse simulations.
We conclude in Sec. \ref{conclusion} by summarizing the results
and possible future directions of study.

\section{Physical System and Analysis of Solution Properties}\label{system}

We are interested in solving the Einstein field equations
\begin{equation}\label{einstein_tens}
R_{\mu\nu} - \frac{1}{2} R g_{\mu\nu} = 8\pi T_{\mu\nu},
\end{equation}
where $R_{\mu\nu}$ is the Ricci tensor, $R\equiv R^\mu{}_\mu$ is
the Ricci scalar, and we use geometric units with
Newton's constant $G$ and the speed of light $c$ set to 1.
We adopt a massless scalar field $\Phi$ as the matter source,
with corresponding stress-energy tensor $T_{\mu\nu}$ given by 
\begin{equation}\label{set}
T_{\mu\nu} = 2 \Phi_{,\mu}\Phi_{,\nu} - g_{\mu\nu} \Phi_{,\gamma}\Phi^{,\gamma},
\end{equation}
and the evolution of $\Phi$ is governed by the wave equation
\begin{equation}\label{phi_eom}
\Box \Phi \equiv \Phi_{;\mu}{}^\mu = 0. 
\end{equation}
Note that (\ref{set}) differs by a factor of $2$ from the convention of 
Hawking and Ellis \cite{hawking_ellis}, which amounts to rescaling $\Phi$ 
by a factor of $\sqrt{2}$.

We restrict our attention to axisymmetric spacetimes without angular momentum,
and choose the following cylindrical coordinate system, adapted to the
symmetry
\begin{eqnarray}
ds^2 = -\alpha^2 dt^2  
       + \psi^4 \bigl[   \,   (d\rho + \brho dt)^2 
\quad\qquad\qquad\qquad
\nonumber\\
                       + (dz + \bz dt)^2 
                       + \rho^2 e^{2 \rho\sigmabar} d\phi^2 
                \bigr].
\label{metric}
\end{eqnarray}
The axial Killing vector is $(\partial/\partial \phi)^\mu$ and hence
all the metric functions $\alpha,\brho,\bz,\psi$ and $\sigmabar$, and
scalar field $\Phi$ depend only on $\rho,z$ and $t$. 

We use the $(2+1)+1$ formalism \cite{maeda} to arrive at 
the system of partial differential equations (PDEs) that we need to solve,
which in the absence of angular momentum is the same set of
PDEs that the ADM decomposition provides. The Hamiltonian constraint
yields an elliptic PDE for the conformal factor $\psi$, and the $\rho$ 
and $z$ momentum constraints give elliptic PDEs for the $\rho$ 
and $z$ components of the shift vector, $\brho$ and $\bz$,  respectively.
We choose maximal slicing, in particular $K_a{}^a=0$, where $K_a{}^b$ is the 
extrinsic curvature tensor of $t={\rm const.}$ slices;
this condition gives an elliptic equation for the lapse function $\alpha$. 
We convert the hyperbolic evolution equations for $\sigmabar$ and $\Phi$ 
to first order form by defining
``conjugate'' variables $\omegabar$ and $\Pi$ by
\begin{equation}\label{omegabar_def}
\bar \Omega \equiv \frac{- 2K_\rho{}^\rho - K_z{}^z}{\rho}
\end{equation}
and
\begin{equation}\label{Pi_def}
\Pi \equiv \frac{\psi^2}{\alpha} \left(\Phi_{,t} - \brho \Phi_{,\rho} 
                                       + \bz \Phi_{,z}\right)
\end{equation}
respectively. 

We thus end up with a mixed hyperbolic-elliptic system of PDES for the 
$8$ variables $\alpha, \psi,\sigmabar,\brho,\bz,\omegabar,\Phi$ and $\Pi$ 
that we approximately solve using
second-order accurate finite difference (FD) techniques. 
The hyperbolic FD equations are solved using an iterative Crank-Nicholson 
scheme with adaptive mesh refinement (AMR), and the elliptic FD 
equations are solved (on the adaptive grid hierarchy) using the FAS 
multigrid algorithm. At $t=0$, we freely specify $\sigmabar,\omegabar,\Phi$
and $\Pi$, then solve the $3$ constraint equations and slicing condition
for the remaining variables. After $t=0$, we continue to use the
momentum constraints to solve for $\brho$ and $\bz$, and the
slicing condition for $\alpha$, but in lieu of the Hamiltonian constraint,
we update $\psi$ using the first order in time evolution equation that 
follows from the definition of the the extrinsic curvature.  (Thus, 
we employ a {\em partially constrained} evolution.)
We add Kreiss-Oliger dissipation to the differenced form of the 
hyperbolic equations, to reduce unwanted (and un-physical) high-frequency 
components in their solutions.
For outer boundary conditions, we apply outgoing radiation (or Sommerfeld)
conditions on $\Phi,\Pi,\sigmabar$ and $\omegabar$, and appropriate
asymptotic fall-off behavior for the remaining variables, assuming an
asymptotically flat coordinate system. 

More details on the boundary conditions,
system of equations and the numerical scheme (including various 
tests) can be found in \cite{graxi_intro}; a detailed description of the 
AMR implementation is given in \cite{pretorius_amr}

\subsection{Analysis of Solution Properties}
In Section~\ref{results} we will quantitatively describe the 
near-critical solution for any given family of initial data 
by measuring its associated scaling exponent ($\gamma$), echoing parameter
($\Delta$), the local minima/maxima attained by the scalar field during
each half-echo, and deviations of the scalar field from a spherically 
symmetric profile. That we {\em can} define such properties for all the
solutions is an indication that they are similar enough that a comparison
is meaningful. However, it is {\em not} a trivial task to 
compute some of these quantities, because we need to make sure that we are 
calculating them in a coordinate independent fashion. Our coordinate system
is not ``symmetry-seeking''~\cite{garfinkle_gundlach}, 
and the initial data is sufficiently
different among the various families that we can expect, and in 
some cases clearly see, ``gauge'' differences between solutions
that are apparently quite similar. 

The simplest quantity to calculate is the scaling exponent
$\gamma$. We use the method proposed by Garfinkle and Duncan 
\cite{garfinkle_duncan}, whereby we measure the maximum value $R_{\rm m}$ 
attained by the absolute value of the 
Ricci scalar, $|R|$, in a set of sub-critical evolutions; 
$\gamma$ can then be obtained from the following property of
near-critical solutions
\begin{equation}\label{lnrp_def}
\ln|R_{\rm m}| \approx -2\gamma \ln\pbar + w(\ln\pbar) + {\rm const.}
\end{equation}
Here $\pbar\equiv \ps-p$ and $w$ is a periodic function of 
its argument with period $\Delta/(2\gamma)$ that describes a 
small ``wiggle'' superimposed on an otherwise linear relationship.
As a result, we can also use (\ref{lnrp_def})
to obtain an estimate for $\Delta$. 
We note that the effectiveness of our use of~(\ref{lnrp_def}) to 
compute $\gamma$ and $\Delta$ is predicated on the degree to which 
our computed near-critical solutions
{\em are} well approximated by a discretely self-similar solution 
with a single unstable mode.  In addition, although~(\ref{lnrp_def}) provides
the only method we use to estimate $\gamma$, we also measure 
$\Delta$ using a more direct procedure outlined below. 

The direct comparison of results from our axisymmetric code to those 
from a spherically symmetric computation presents more of a challenge.
In order to compare ``local self-similar solutions''---portions of 
the computed spacetime that appear to be approximately self-similar about 
some center of symmetry---we need an invariant way of slicing the spacetime 
in the region of interest.  To accomplish this, we use a sequence of 
outgoing null hypersurfaces, 
starting from the local center of symmetry $(\rho,z)=(0,z_0)$,
to generate the common slices along which
we compare $\Phi$. To construct each such null
hypersurface, we evolve a family of null geodesics,
with affine parameter $x$ and initial tangent vectors 
equally spaced in $\theta\equiv\tan^{-1}(\rho/(z-z_0))$, outward
from $(\rho,z) = (0,z_0)$.  The geodesics are synchronized by
setting $dx/d\tau=1$ at the start of integration, where   
$\tau$ is the proper time measured by a timelike observer
that is stationary relative to the center of symmetry. $\tau$
is the time of relevance to critical collapse, for in
coordinates $\ln(\taus-\tau)$, where $\taus$ is the 
{\em accumulation} point of the critical solution (i.e. the central 
proper time of the central singularity formed by the cascade of 
the critical solution down to infinitesimally small scales),
the central value of the scalar field is a periodic function
of $\ln(\taus-\tau)$, with period $\Delta$. 
Estimation of the period of the profile of $\Phi$ along 
the local center of symmetry, with respect to $\tau$ thus gives 
us the alternate method for computing $\Delta$.

During a simulation, we integrate $x$ as a function of $t$
for each null geodesic labeled by $\theta_0\equiv\theta(x=0)$, 
and record $\Phi(x,\theta_0)$ (we typically use $50$ geodesics
per slice, linearly spaced in $\theta_0$).
If two solutions from different
families of initial data do locally tend to the same 
discretely self-similar solution, then $\Phi(x,\theta_0)$
(synchronized so that the null integration
is started at the same time within the periodic oscillation) 
will tend to the same function, regardless of differences
in the $(\rho,z,t)$ coordinate systems between the two solutions.

As a final comment in regards to our analysis, we note that to calculate
$\tau$ we integrate a 
central timelike geodesic, and measure proper time along it. 
We can do this for families of initial 
data that are symmetric about $z=0$, for then we know that
the center of symmetry will, at least initially, be at $(0,0)$. 
An interesting aspect of the numerical solution is that
truncation error effects cause a small drift to occur in the 
$z$ location of the local center of symmetry, during a 
near-critical evolution. This drift is quite small (and does appear 
to converge away with increasing resolution), 
typically being less than 1 part in $10^6$
of the size of the computational domain. However, because of
the exponentially decreasing length scales that arise in a critical
collapse, this is a {\em huge} drift relative to the size of the
local self-similar region at late times. Hence, if we simply
measured central proper time at $(\rho,z)=(0,0)$, and correspondingly
integrated null geodesics from this location, we would entirely
miss the relevant part of the solution. Fortunately,
the timelike observer initially placed at $(0,0)$ experiences
an identical drift, and so we can use its location and proper
time to do the desired measurements. For initial data
that is not plane-symmetric (the only such family described
in the next section is the ``anti-symmetric'' example) we have not yet been
able to devise a method to accurately track the local center
of symmetry for long periods of time, and hence have not been
able to calculate $\tau$ for these families. However, 
at least at key moments during the evolution, we are able
to accurately determine the center of symmetry (by looking at local minima or
maxima of $\Phi$, for instance), to use as the starting point for the null integration.

\section{Results}\label{results}

Here we present results from the critical collapse of several families of
initial data. These families consist of a time-symmetric series of prolate spheroids, with 
ellipticity $\epsilon$ (defined by looking at surfaces of constant $\Phi$):
\begin{eqnarray}
\Phi(0,\rho,z)&=&A e^{-\left(\rho^2+(1-\epsilon^2) z^2\right)}, \nonumber \\
\Pi(0,\rho,z)&=&0, 
\label{prolate_ic}
\end{eqnarray}
and an initially ingoing distribution in $\Phi$ that is
anti-symmetric about $z=0$, i.e. $\Phi(t,\rho,z)=-\Phi(t,\rho,-z)$:
\begin{eqnarray}
\Phi(0,\rho,z)&=&A z e^{-\left(\sqrt{\rho^2+z^2}-R_0\right)^2}, \nonumber \\
\Pi(0,\rho,z)&=&-\Phi(0,\rho,z). 
\label{anti_ic}
\end{eqnarray}
In all cases we set $\sigmabar(0,\rho,z)=0$ and $\omegabar(0,\rho,z)=0$,
and vary the amplitude $A$ when searching for the threshold solution.
We show results for six families of (\ref{prolate_ic}), with
$\epsilon^2=0,1/3,1/2,2/3,3/4$ and $5/6$. 
In (\ref{anti_ic}), $R_0$ is a parameter describing how far the initial pulse
of matter begins from the origin; we have chosen $R_0=3$. In all
cases presented here the outer boundary of the computational
domain is at $\rho=|z|=10$, though in other simulations we have 
varied its position to make sure that the above choice 
does not significantly impact the results.
The base level in the adaptive hierarchy used a resolution of 
$65 \times 129$ points, and up to 28 additional 2:1-refined levels 
were used in the most nearly critical case.  Our AMR implementation
is based on the algorithm of Berger and Oliger~\cite{berger_oliger}, 
wherein regridding is determined through estimates of the local truncation
error (solution error) in the computed solution.  The key control parameter
that determines placement of refinements is the truncation error threshold,
$\tau_{\rm m}$: mesh refinements are introduced in an attempt to keep 
the magnitude of the local truncation error estimate $\le \tau_{\rm m}$
throughout the solution domain.
For each family of initial data studied, we 
generally tuned to threshold using three different values of
$\tau_{\rm m}$, namely $\tau_{{\rm m}0}$, $\tau_{{\rm m}0}/2$
and $\tau_{{\rm m}0}/4$. Most of the data presented here are from 
$\tau_{\rm m}=\tau_{{\rm m}0}/4$ runs (i.e. finest effective resolution),
with the results computed using the less stringent values of $\tau_{\rm m}$ 
then being used to give some estimate of how
close to the continuum solution we may be (though convergence testing
with an adaptive code is not trivial, particularly in the critical limit). 

\begin{figure}
\begin{center}
\includegraphics[width=8.5cm,clip=true,draft=false]{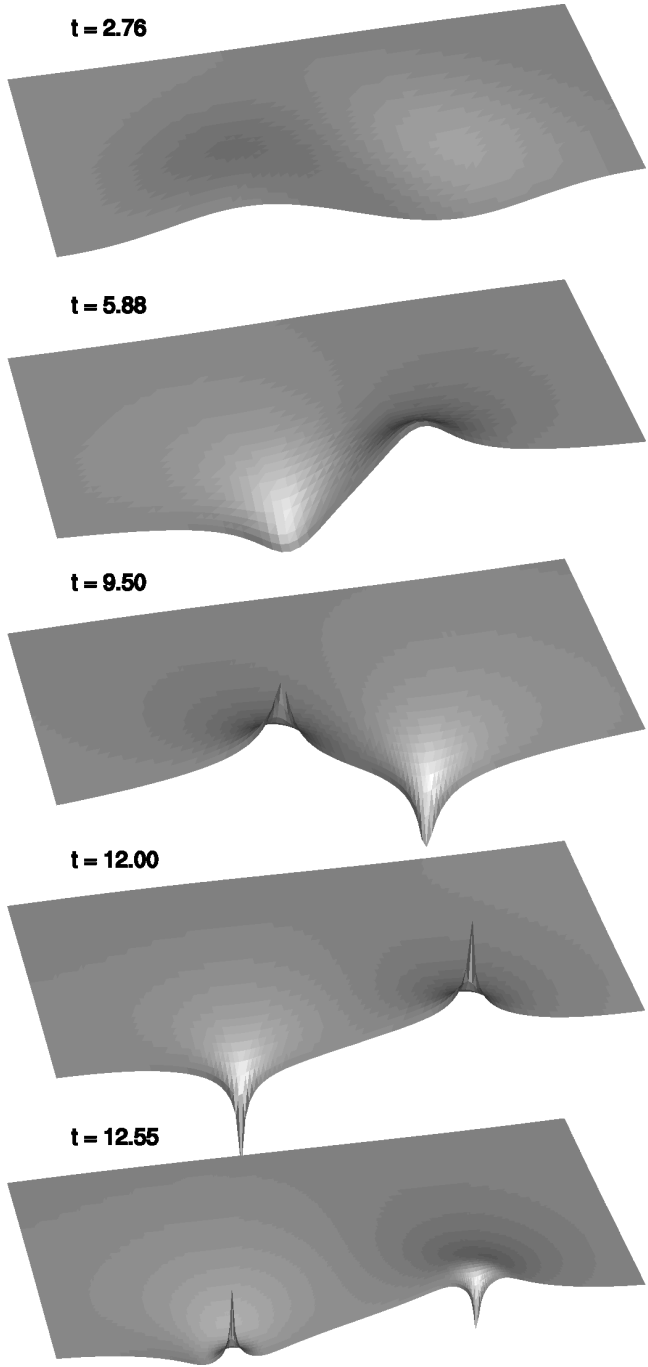}
\end{center}
\caption
{Several frames of $\Phi(\rho,z,t)$ from the evolution of near-critical, anti-symmetric
initial data (\ref{anti_ic}). The figures span the first several
half-echoes of the local self-similar solutions, and the particular times shown
correspond to when the scalar field reaches a local minima/maxima. 
The height of each surface
represents the magnitude of $\Phi$, and the coordinate domain of each figure is
$[0 .. 2.5,-2.5 .. 2.5]$ in $[\rho,z]$ (the axis $\rho=0$ is the nearest edge
of each plot, and positive to negative $z$ runs from left to right). }
\label{asym_ex}
\end{figure}

There are of course, infinitely many different parametrized families that 
we could have considered---those used to generate the results discussed 
here were chosen for the following specific reasons.
First, the anti-symmetric configuration (\ref{anti_ic})
provides a more drastic departure from spherical symmetry than
any family of data that can smoothly be deformed into a spherical distribution
(such as (\ref{prolate_ic}) by letting $\epsilon\rightarrow 0$). 
In this regard we note 
that one of the characteristic features of spherical scalar field critical 
collapse
is that the ``central'' value of $\Phi$ oscillates between specific 
extremal values $\pm \Phi_0$; clearly the anti-symmetric property of 
(\ref{anti_ic}) allows no such oscillation. One might therefore 
expect that evolutions with this type of initial data might produce
a qualitatively different critical solution than the spherically
symmetric one.  However, as shown in Figure \ref{asym_ex}, at threshold 
{\em two} spherical-like echoing solutions develop off-center at $z=\pm
z_c(t)$.

\begin{figure}
\begin{center}
\includegraphics[width=7.0cm,clip=true,draft=false]{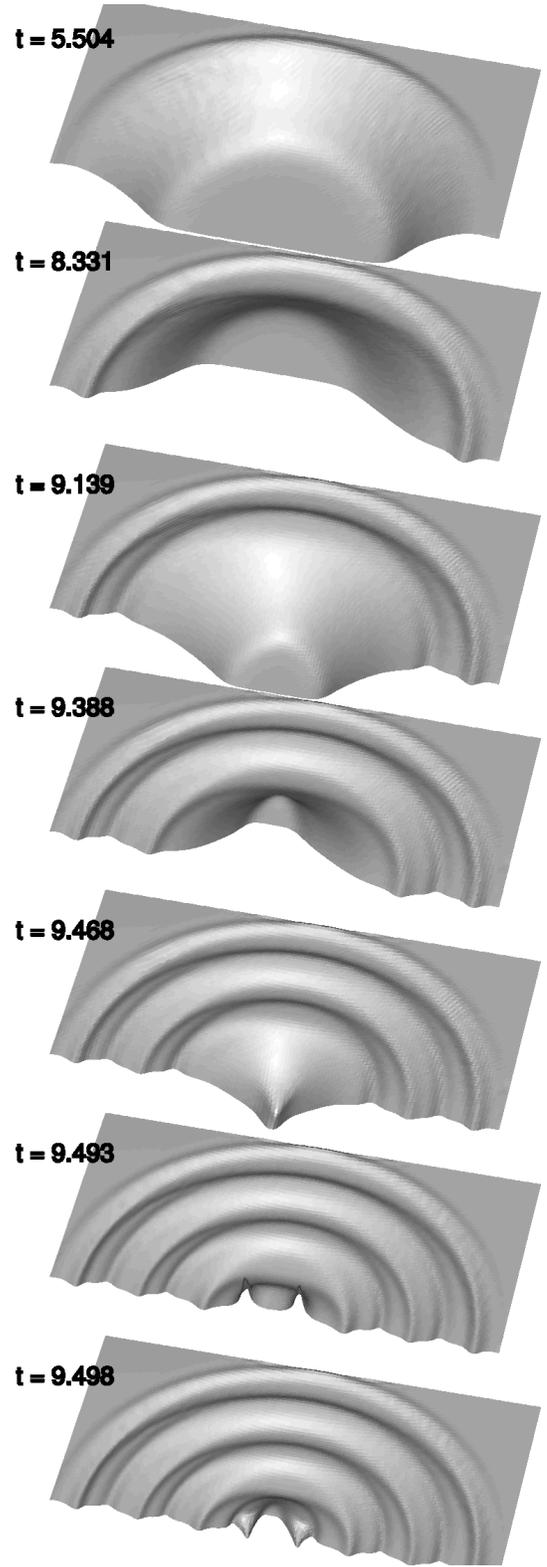}
\end{center}
\caption
{Several frames of $\Phi(\bar{r},\theta,t)$ from the evolution of near-critical,
$\epsilon^2=3/4$ prolate initial data (\ref{prolate_ic}). Here we have
transformed to coordinates $\bar{r}=\ln(\sqrt{\rho^2+z^2} + e_0)-\ln e_0$ (with
$e_0=2$x$10^{-4}$) and $\tan\theta=\rho/z$, to give a better view of the 
initial self-similar nature of the solution. $[\bar{r},\theta]$ ranges from
$[0..\approx 10.8,0..\pi]$, with the axis $\rho=0$ being the nearest edge
in each figure. The height of each surface represents the magnitude of $\Phi$.
The times shown correspond to
the times when $\Phi$ reaches a local minima/maxima, demonstrating the
bifurcation that occurs after about $2$ self-similar echoes of the field.}
\label{prolate_1_4_ex}
\end{figure}

Second, we include the prolate family because the initial amplitude of 
the putative second unstable mode seems to be closely related to 
the prolateness of the initial distribution (rather, for instance, 
than asymmetries in $\Phi$ within an 
imploding spherical shell). Therefore, the parameter $\epsilon$
in (\ref{prolate_ic}) allows us to demonstrate the effect of adding more
(larger $\epsilon$)
or less of the unstable mode. The axisymmetric instability, once
it has grown beyond a certain amplitude, causes
a near-spherical threshold solution to ``bifurcate'' into two
echoing solutions, separated by some distance along the axis. As an example, 
Figure \ref{prolate_1_4_ex} shows several time-instants from the near-critical 
evolution of initial data with $\epsilon^2=3/4$, transformed to logarithmic 
coordinates in space to better illustrate the self-similar nature of the 
initial critical behavior. Note that this bifurcation is qualitatively 
different from 
the two echoing solutions observed in anti-symmetric collapse---there, by 
construction, no self-similar behavior is seen about 
$z=0$, and there are two (out of phase) echoing solutions from the beginning. 
Furthermore, the initial separation of the two (in phase) echoing solutions 
arising from a bifurcation is related to the smallest length scale that 
developed in the single, origin-centered echoer prior to the 
bifurcation.  In contrast, the
separation of the two anti-symmetric echoing solutions is related to a length 
scale in the initial data.  Moreover, if there really is a second unstable
mode, then each
of the anti-symmetric echoers should 
also be subject to that instability and eventually bifurcate, and we do see 
some evidence for this. 

With double precision arithmetic, we are able to tune
the initial amplitude of a given family to within a part in $10^{15}$ 
of threshold\footnote{Due to lack of computational resources (memory and time)
we were only able to tune the $\epsilon^2=3/4$ and
$\epsilon^2=5/6$ cases to within roughly a part in $10^{12}$}, corresponding
to about 3 full echoes of the spherically symmetric critical solution. The
growth of the instability is sufficiently small that after 
3 echoes we do not yet see a bifurcation for $\epsilon^2\leq2/3$; for
$\epsilon^2=3/4$ and $\epsilon^2=5/6$ we see a bifurcation after approximately
$2$ and $1$ $1/2$ echoes respectively.

\begin{table}
\begin{center}
\begin{tabular}[t]{| c || c | c | c | c |}
\hline
$\epsilon^2$ & $\gamma$ & $\langle\Delta\rangle^1$ & $\langle\Delta\rangle^2$ &
$\langle|\Phi_c|\rangle$  \\
\hline
\hline
$0$       & $0.382\pm2\%$   & $3.44\pm1\%$  & $3.49\pm3\%$ & $0.431\pm2\%$ \\
$1/3$     & $0.380\pm2\%$   & $3.41\pm1\%$  & $3.43\pm3\%$ & $0.431\pm2\%$ \\
$1/2$     & $0.375\pm3\%$   & $3.37\pm1\%$  & $3.39\pm4\%$ & $0.430\pm2\%$ \\
$2/3$     & $0.346\pm3\%$   & $3.13\pm1\%$  & $3.08\pm4\%$ & $0.419\pm3\%$ \\
$3/4 (a)$ & $0.313\pm4\%$   & $2.87\pm4\%$  & $3.03\pm5\%$ & $0.396\pm6\%$ \\
$3/4 (b)$ & $0.40\pm10\%$   & ---           & $\approx 3$  & $0.40\pm8\% $ \\
$5/6 (a)$ & $0.28\pm10\%$   & $\approx 2$   & $\approx 1$  & $0.36\pm7\% $ \\
$5/6 (b)$ & $0.41\pm10\%$   & ---           & $\approx 3$  & $0.36\pm10\%$ \\
\hline
\hline
{\em AS}  & $0.383\pm2\%$   & ---           & $3.49\pm3\%$ & $0.434\pm3\%$ \\
\hline
\end{tabular}
\end{center}
\caption
{Critical parameters of the prolate families (\ref{prolate_ic}) and 
the anti-symmetric family ($AS$) (\ref{anti_ic}). 
$\gamma$ is obtained from a least-squares fit to the data shown in 
Figure \ref{lnpr}, $\langle\Delta\rangle^1$ is the average
value of $\Delta$ measured between adjacent extremes in $\Phi_c$ as shown 
in Figure \ref{cent_phi} (only using data from intermediate times),
$\langle\Delta\rangle^2$ is the average value of $\Delta$ 
inferred from the periodic oscillations in Figure \ref{lnpr},
and $\langle|\Phi_c|\rangle$ is the average absolute value of
the extremes of $\Phi_c$ in Figure \ref{cent_phi} 
(again using data from intermediate times). For the two prolate cases that 
bifurcate---$\epsilon^2=3/4$ and $\epsilon^2=5/6$---we list estimates
of these parameters (where possible) before (a) and after (b) the bifurcation.
For the anti-symmetric case, we do not have data for $\Phi_c$ versus
central proper time; $\langle|\Phi_c|\rangle$ in that case is calculated
as half the average {\em difference} between subsequent local extremes
in $\Phi(\rho=0,z,t)$ about one of the local self-similar solutions.
See the text for a discussion on how the estimated 
uncertainties were calculated.
\label{tab_crit_exp}}
\end{table}

\begin{figure}
\begin{center}
\includegraphics[width=8.5cm,clip=true,draft=false]{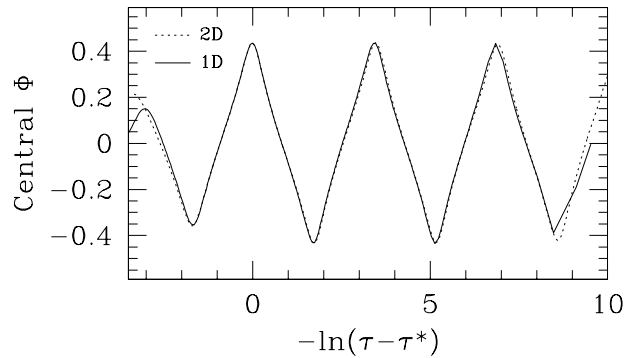}
\end{center}
\caption
{Comparison between results from our 2D axisymmetric 
code ($\epsilon=0$ initial
data) and a 1D spherically symmetric code \cite{choptuik}. Plotted is the
central value of $\Phi$ as a function of logarithmic central proper
time, for the most nearly critical simulations obtained in either case.}
\label{cent_phi_1d}
\end{figure}

\begin{figure}
\begin{center}
\includegraphics[width=18cm,clip=true,draft=false]{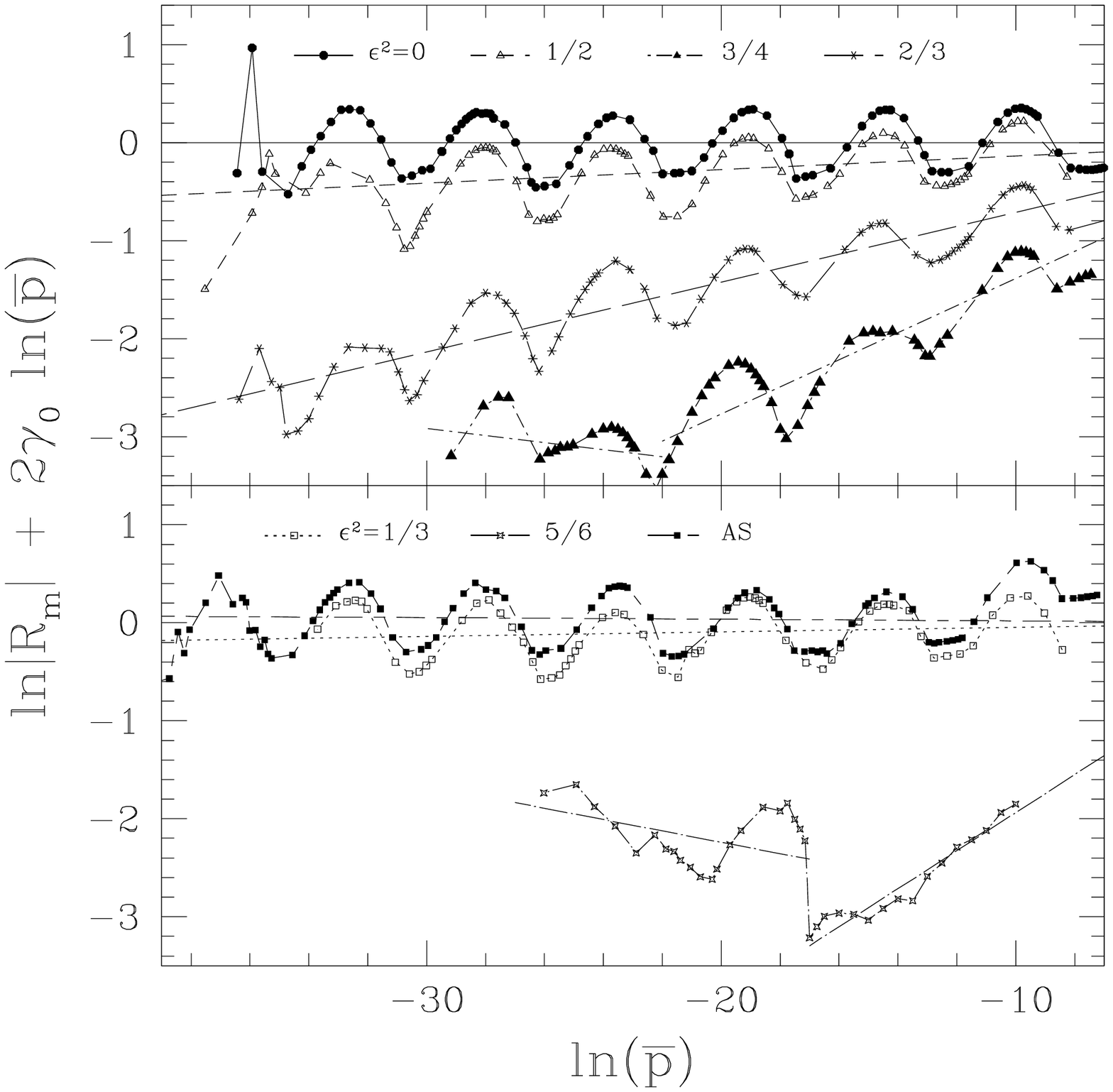}
\end{center}
\caption
{Plots of the logarithm of the maximum absolute value $|R_{\rm m}|$ 
attained by the
Ricci scalar, during sub-critical evolution, vs. the logarithm
of the distance in parameter space $\bar{p}=p^\star-p$ from the
(estimated) critical parameter $p^\star$, for all families of initial
data considered (using the lowest $\tau_m$ data). To avoid clutter
in the figure, we have placed the data from each family on one
of two identical panels. To facilitate comparison,
the line $-2\gamma_0\ln\bar{p}$ has been subtracted from each curve,
with $\gamma_0=0.382$ (the estimated value from the $\epsilon=0$ family).
Also, the intercept of each curve has been set to $\bar{p}=0$.
The estimated linear relationships used to calculate $\gamma$ in 
Table \ref{tab_crit_exp} are also shown in the figure. 
}
\label{lnpr}
\end{figure}

\begin{figure}
\begin{center}
\includegraphics[width=18cm,clip=true,draft=false]{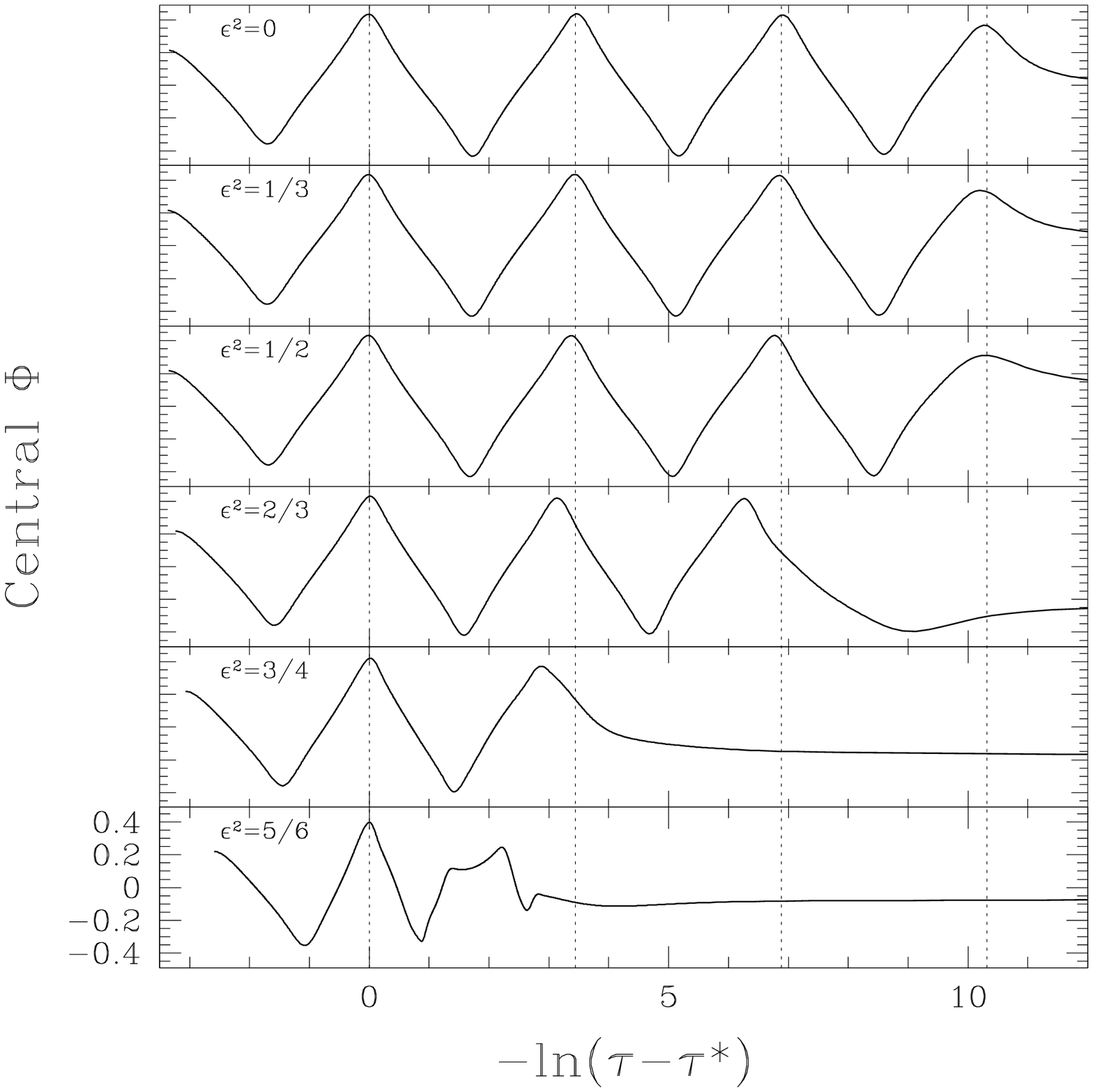}
\end{center}
\caption
{The central value of $\Phi$ as a function of logarithmic central proper
time, from the most nearly critical simulations of the prolate
collapse families. The horizontal axis for each family was shifted
so that the first maximum of each curve occurs at $-\ln(\tau-\taus)=0$.
$\taus$ is the accumulation point of each family (prior
to the bifurcation for the $\epsilon^2=3/4$ and $\epsilon^2=5/6$ cases),
calculated by {\em assuming} that the intermediate time behavior is 
discretely self-similar, and then finding the $\taus$ for each case that 
minimizes the variance in $\Delta$
computed between pairs of adjacent minima/maxima in $\Phi_c$.
To aid in comparison, dashed vertical lines have been drawn at intervals
of $3.44$ in $\tau$, which is the estimated spherical echoing exponent.}
\label{cent_phi}
\end{figure}

Table \ref{tab_crit_exp} summarizes measurements made of the critical
parameters---namely $\gamma,\Delta$ and the amplitude of each echo in
$\Phi$---from the $\tau_{m0}/4$ simulations for each family of initial
data (except for the $\epsilon^2=5/6$ case, where the increasing computational
demands, resulting from larger, more elongated grids that are produced in the 
hierarchy for higher values of $\epsilon$, prevented us from computing 
with anything but 
$\tau_m=\tau_{m0}$). For the two simulations with the largest values
of $\epsilon$,  we list
parameters obtained before and after the bifurcation, where possible. As with the
anti-symmetric case, our method of geodesic integration cannot track moving
centers, and so we cannot provide a direct estimate of $\Delta$ after a
bifurcation. Also,
for the $\epsilon^2=5/6$ case, we do not see a very
distinctive periodic oscillation in the $\ln R_{\rm m}$ vs. $\ln\bar{p}$ plot,
and thus can only provide a rough guess for $\Delta$ from that information.
Most of the data in this table was gathered from Figures \ref{lnpr} and \ref{cent_phi},
which show $\ln R_{\rm m}$ vs. $\ln\bar{p}$ (with the linear relationship from
the spherical family subtracted to better differentiate the plots) and 
$\Phi_c$, the central
value of $\Phi$, vs. logarithmic central proper time for the prolate
families prior to bifurcation, respectively. Also, Figure \ref{cent_phi_1d}
shows the same type data displayed in Figure \ref{cent_phi}, but for the case 
$\epsilon=0$, and with the addition of an 
overlay of data obtained with a spherically symmetric 1D code
\cite{choptuik}.  The good agreement between the results from the 
axisymmetric and spherical computations provides 
a measure of confidence in the correctness and level of accuracy
of our 2D code.

The quoted uncertainty of a given value in Table \ref{tab_crit_exp} was 
calculated as the sum of the estimated truncation error from a convergence 
calculation using the different $\tau_m$ runs, 
and the standard deviation from the relevant 
averaging/fitting operation (except for $\epsilon^2=5/6$, where we could not 
estimate the truncation error as
we only have data from a single value of $\tau_m$). However, in a sense
these uncertainties are ``optimistic,'' for we have {\em not}
accounted for possible systematic errors. Chief among these (in particular
away from spherical symmetry) are the assumptions of discrete self-similarity,
which was used to define $\taus$ in Fig.\ref{cent_phi}, and the assumption
that the linear and periodic parts of Fig.\ref{lnpr} are directly related
to $\gamma$ and $\Delta$ respectively. For several of the simulations we
have checked that the following numerical parameters are {\em not} significant
sources of systematic error: outer boundary location, Dirichlet vs. Neumann
conditions on $\alpha, \brho$ and $\bz$ at the outer boundary, and free
vs. constrained evolution for $\psi$.

\begin{figure}
\begin{center}
\includegraphics[width=8.75cm,clip=true,draft=false]{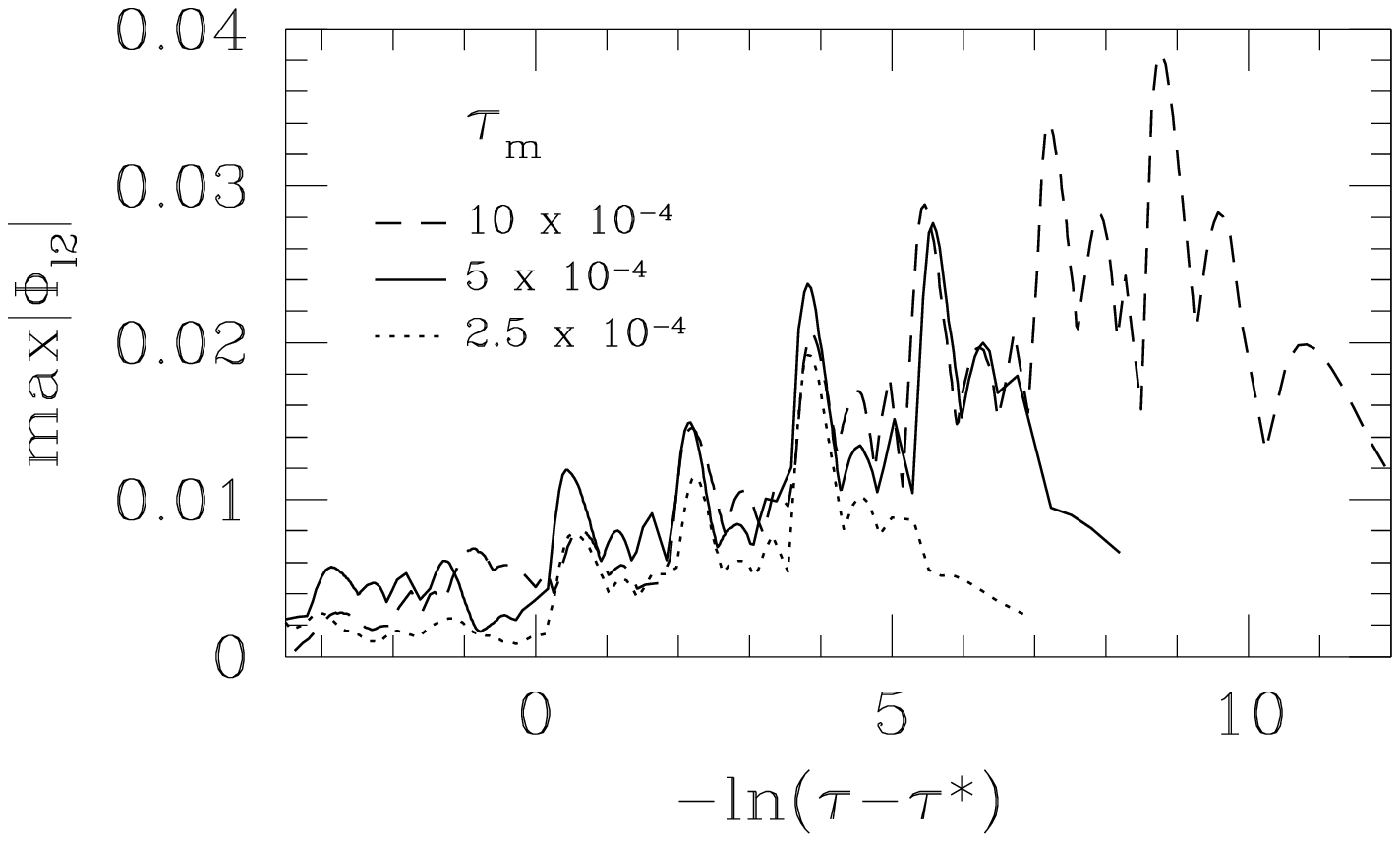}
\end{center}
\caption
{The maximum absolute value of the $\ell=2$ ($m=0$) spherical harmonic
component of $\Phi$, $\Phi_{\ell2}$, in near-critical $\epsilon=0$ collapse, 
measured along outgoing null slices of the spacetime, from simulations with 
3 different values of the maximum truncation error estimate $\tau_m$. 
The graph indicates similar growth rates independent of the effective
resolution. In order to better show the similarity of the growth rates,
the data for the higher truncation error thresholds have been shifted
along the horizontal axis which labels logarithmic central proper time
(about $-3.4$ for the $\tau_0/2$ 
case and $-5.1$ for the $\tau_0/4$ data).
Thus, at similar un-shifted times, the graph indicates that
the amplitude of $\Phi_{\ell2}$ decreases with $\tau_m$, as expected.}
\label{MA_ssb}
\end{figure}

Note that our coordinate
system is {\em not} adapted to spherical symmetry, and during an
evolution of $\epsilon=0$ initial data, spherical symmetry is only preserved
to within an amount proportional to the truncation error, and so
will eventually exhibit the apparent second growing mode. In a certain 
sense this
is a desirable feature, for at late times during an $\epsilon=0$ evolution
this mode is the only one (apart from the unstable spherical mode) that should 
be visible perturbing the spherical
solution; an $\epsilon>0$ evolution exhibits a host of additional, 
decaying asymmetric modes that prevent us from easily measuring the properties
of the non-spherical growing mode. 
To this end, in Figure \ref{MA_ssb} we show plots of the maximum absolute value
of the $\ell=2$ ($m=0$) spherical harmonic
component of $\Phi$, denoted $\Phi_{\ell2}$, 
in near-critical $\epsilon=0$ collapse, as
measured along outgoing null slices of the spacetime (in other words,
we decompose $\Phi(x,\theta_0)$, constructed as described in Sec. \ref{system},
into its spectral coefficients for each $x$---the 
$\ell=2$ component is $\Phi_{\ell2}$.).
We show results from
simulations with 3 different values of the maximum truncation error estimate $\tau_m$,
demonstrating the expected behavior that $\Phi_{\ell2}\rightarrow 0$ in the
limit $\tau_m\rightarrow 0$.

\onecolumngrid

~\\
\hrule 
~\\

\vspace{0.2in}

\twocolumngrid

We now argue that Fig. \ref{MA_ssb} also gives some evidence that the 
instability we do see in the numerical solution may be an actual feature of
the continuum solution, and not a truncation-error-driven phenomenon. 

Assume that the numerical solution has a well-behaved Richardson expansion.
Then we expect any well-defined continuum property of the solution, 
such as the growth rate, $\lambda_i$, of a perturbative mode 
to have a similar expansion:
\begin{equation}\label{lambda_exp}
\hat{\lambda}_i = \lambda_i + h f(\vec{x}) + O(h^2),
\end{equation}
where $\hat{\lambda}_i$ is the numerically measured growth rate, $h$ is the
discretization scale (we have assumed a first order accurate discretization), 
and $f(\vec{x})$ is some function of the continuum
solution variables $\vec{x}$. Of course, in an adaptive scheme there is no single
scale $h$; however, individual grids within the hierarchy do admit Richardson
expansions, and hence we can loosely think of (\ref{lambda_exp}) holding over the
hierarchy with some effective $h$ that would be related to the
maximum truncation error estimate $\tau_m$. In \cite{martin_garcia_gundlach},
the real part of the largest eigenvalue of any non-spherical mode 
perturbing the critical solution was found to be $\lambda_2\approx -0.02$;
i.e. a decaying mode, and the corresponding eigenfunction had the angular 
dependence of the $\ell=2,m=0$ spherical harmonic $Y^\ell_m$.
This magnitude of decay is about 100 times smaller 
than the growth rate of the dominant spherically symmetric mode, 
that has $\lambda=1/\gamma\approx 2.7$. Thus, looking at (\ref{lambda_exp}),
it is certainly plausible that in a numerical scheme, even if $h$ were
small enough to reasonably accurately model the dominant feature of a 
solution (as we evidently are from the comparison in Figure \ref{cent_phi_1d}), 
it might still be large enough to significantly affect sub-dominant
features of the solution, such as a small $\lambda_i$ in (\ref{lambda_exp}).

\begin{figure}
\begin{center}
\includegraphics[width=8.75cm,clip=true,draft=false]{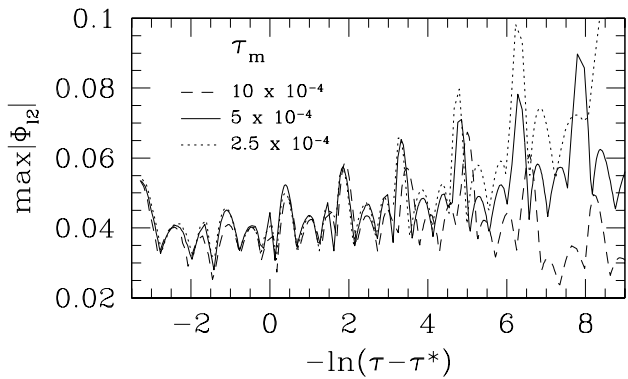}
\end{center}
\caption
{The maximum absolute value of the $\ell=2$ ($m=0$) spherical harmonic
component of $\Phi$, $\Phi_{\ell2}$, in near-critical $\epsilon^2=2/3$ 
collapse, measured along 
outgoing null slices of the spacetime, from simulations with 3 different 
values of the maximum truncation error estimate $\tau_m$. 
The horizontal axis labels the logarithmic central
proper time when the given outgoing null surface intersects the origin.
Note the different vertical scales when comparing this plot to the similar 
one for $\epsilon=0$ in Figure \ref{MA_ssb} (and we have {\em not} shifted the
data here).}
\label{MA_1_3}
\end{figure}

We should then be able to see
a significant effect when changing $h$; however, in Figure \ref{MA_ssb}, 
even though the initial amplitude of the asymmetry decreases as $\tau_m$
decreases (as expected), the apparent growth rate that we obtain, namely
$\lambda_2 \approx 0.1 - 0.4$, does not noticeably
change within the relatively large uncertainty of the measurement\footnote{We
estimate $\lambda_2$ by assuming exponential growth between adjacent, local
maxima of the plots in Figure \ref{MA_ssb} (and \ref{MA_1_3} later on), and then
averaging the growth rate found over the number of half-echoes where the 
growth is clearly visible (so unfortunately with $\epsilon=0$ data, the more
accurate the solution with smaller $\tau_m$, the fewer data points we
have to estimate the growth).}. On the other hand, we may still be too far
from the convergent regime to measure $\lambda_2$ (so that higher order terms
in (\ref{lambda_exp}) are still important). Note that we also cannot
conclusively say that the growing mode we see has a pure $\ell=2$ angular 
dependence, but it appears that 
the $\ell=2$ mode is at least an order of magnitude larger
than any of the other asymmetric modes we find in the spectral decomposition.
However, it must be noted that for computations with any of the three 
values of $\tau_m$ adopted  we use 50 points in $\theta$
along which we integrate null curves, and so do not have good accuracy for 
determining the higher $\ell$ modes.

In Figure \ref{MA_1_3} below we show a plot of the growth 
rate of $\Phi_{\ell2}$ from $\epsilon^2=2/3$ near critical solutions.
This value of $\epsilon$ is the smallest, {\em non-zero} value considered that
clearly shows growth of the asymmetry during the roughly
three self-similar echoes of evolution; in the $\epsilon^2=1/3$ and
$\epsilon^2=1/2$ cases, early-time evolution of the $\ell=2$ spectral 
component is dominated by decaying modes. 
For the $\epsilon^2=2/3$ data in 
Figure \ref{MA_1_3} we apparently {\em are} converging to the growth shown; 
i.e. as was the case for the spherically symmetric initial data, the 
growth does not appear to be 
truncation error dominated.
Estimates of the growth rate from the simulation with the smallest value of 
$\tau_m$ gives $\lambda_2 \approx 0.05 - 0.15$. However, this is 
quite a rough estimate as we cannot disentangle the supposed growing mode from 
the full spectrum of $\ell=2$ modes contributing to the plot shown in 
Figure \ref{MA_1_3}.

Although we appear to be converging to a growth of the asymmetry
in the $\epsilon^2=2/3$ case, and to a bifurcation for the $\epsilon^2=3/4$ case, 
this does not necessarily
prove that the spherically symmetric critical solution has a second unstable mode.
These families are sufficiently aspherical that one can imagine
that the bifurcation is due to some artifact of the initial data---in 
particular
a ``focusing'' effect, as the wave front of an imploding, prolate
distribution of the scalar field will tend to focus to two locations 
on the axis, above and below the origin. If this is the case though, it is
rather surprising that we see self-similar collapse occur about a single
center {\em prior} to the bifurcation. 

\begin{figure}
\begin{center}
\includegraphics[width=8.75cm,clip=true,draft=false]{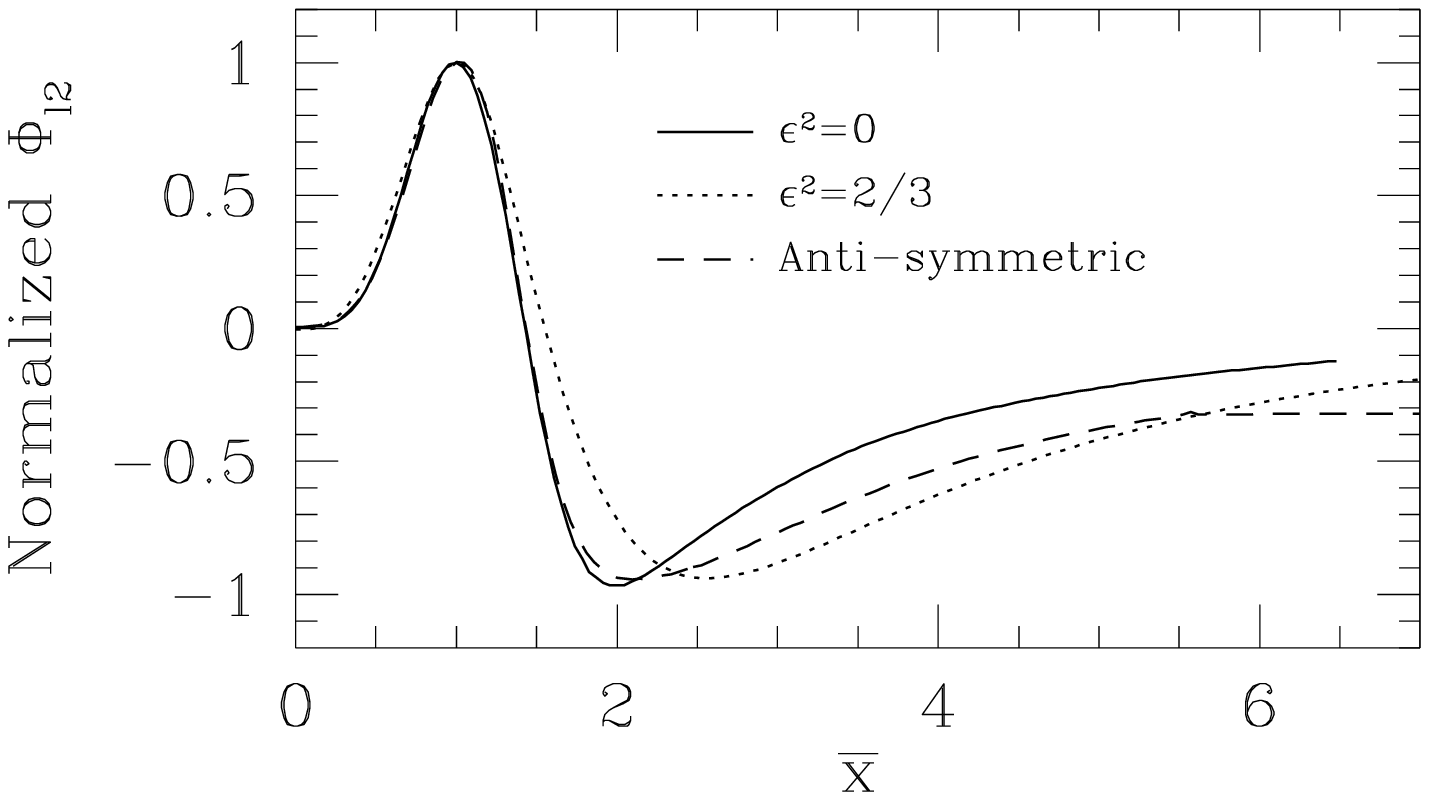}
\end{center}
\caption
{The normalized $\ell=2$ ($m=0$) spherical harmonic component of 
$\Phi$, measured along an outgoing null geodesic starting from
similar times within a (chosen) self-similar oscillation of the
$\epsilon^2=0,\epsilon^2=2/3$ and anti-symmetric near critical
solutions (from $\tau_m=\tau_{m0}/4$ simulations). To facilitate
comparison, we have rescaled the amplitude of each curve so
that the maximum is 1, and rescaled the affine parameter (labeled
$\bar{x}$) along the null curves so that the first maxima of each
is at $\bar{x}=1$. }
\label{L2_comp}
\end{figure}

Finally, in Figure \ref{L2_comp}, we show comparisons of the ``radial''
profile of the $\ell=2$ spherical harmonic component of
$\Phi$, measured along an outgoing null geodesic at (approximately) the 
same time within a self-similar echo\footnote{Roughly one-third of an echo
from the time when the scalar field attains a local maximum (the fourth
such maxima for the $\epsilon^2=0$ and anti-symmetric cases, and
the second maxima for the $\epsilon^2=2/3$ case---see Figure \ref{cent_phi})}, for
the $\epsilon^2=0, \epsilon^2=2/3$ 
and anti-symmetric near-critical solutions (with $\tau_m=\tau_{m0}/4$). 
In the plot, the overall amplitude and affine distance along 
each null curve is rescaled so that the maximum amplitude is one, and
occurs at one in rescaled affine time $\bar{x}$. That the curves from these
representative families do approximately agree provides additional 
evidence that we are
seeing a {\em unique}, asymmetric unstable mode.

\section{Conclusion}\label{conclusion}
We have presented results from a first study of scalar field critical collapse
in axisymmetry in the fully non-linear regime. We find that critical phenomena
is observed at the threshold of gravitational collapse of several families of
asymmetric initial data. The critical solution that unfolds at threshold
can (locally) be described as the spherically symmetric critical solution found in 
\cite{choptuik}, with asymmetric perturbations. However, in contrast to the results of 
\cite{martin_garcia_gundlach}, we find some evidence that a single $\ell=2$
spherical harmonic perturbation does {\em not} decay with time; rather it grows
at a rate of roughly $1/10$ the magnitude of the dominant, spherically
symmetric unstable mode. The nature of this second unstable mode is such that
it causes a self-similar threshold solution, with some asymmetry in it, to
eventually bifurcate into two local, self similar solutions that again
resemble the spherical threshold spacetime. If this second instability is
indeed a property of the spherically symmetric critical solution, then
presumably one (or both if the initial data has reflection symmetry) of the 
new self-similar solutions would bifurcate again,
and so on, resulting in an infinite, ``random walk'' of bifurcations
on ever decreasing scales. Thus the second instability would not completely
destroy the universal nature of generic (axisymmetric) critical collapse,
but rather would alter it in an intriguing, family dependent manner.

To conclusively answer 1) whether there {\em is} a second unstable mode, and
if so 2) how the bifurcation ultimately affects the threshold solution,
is beyond the capabilities of our current code. First, we are using double
precision arithmetic, and this prohibits us from tuning closer than $1$ part in $10^{15}$
of the threshold. Because the echoing exponent of the spherically symmetric solution is
(relatively speaking) so large, $1$ part in $10^{15}$ can only give us about 
three, complete self-similar echoes. This is far from ideal when trying to
estimate the growth (or decay) rate of a mode that may have an e-folding time
on the order of 10-20 echoes. Second, we would like to achieve higher accuracy
than we have been able to attain so far. To do this with the current
code will require that we parallelize it because we have already reached
the practical limits (in terms of memory usage and runtime) imposed by the
hardware to which we have access. Alternatively, one could write a code adapted
to the spherical critical solution (for instance using spherical polar
coordinates with a logarithmic radial coordinate). This would allow one to
obtain greater resolution, with a given amount of resources, than what
we can achieve with our more general purpose cylindrical coordinates. Of
course, spherical polar coordinates would not be well suited to following
a solution beyond a bifurcation, but it should be adequate to study
the growth or decay of perturbations.\\

\section*{Acknowledgments}

The authors would like to thank Kristin Schleich, Bill Unruh and Jim Varah
for valuable conversations.
The authors gratefully acknowledge research support from 
CIAR, NSERC,  NSF PHY-9900644, NSF PHY-0099568, NSF PHY-0139782, 
NSF PHY-0139980, Southampton College, the Izaak Walton Killam Fund and 
Caltech's Richard Chase Tolman Fund.
The simulations shown here were performed on UBC's {\bf vn}
cluster, (supported by CFI and BCKDF), and the {\bf MACI} cluster
at the University of Calgary (supported by CFI and ASRA).

\end{document}